# Periodic structure of spin-transfer current in ferromagnetic multilayers

Pavel Levin
*Mechanical Engineering Department, The City College/CUNY*
*1270 E. 19$^{th}$ St., #2M, Brooklyn, NY 11230*

We show that the drift-diffusion mechanism in a normal-metal layer in combination with the resonance electron-magnon interactions at ferromagnet-normal interface of F-N-F heterostucture creates spatial instability modes and, out of these modes, a quasi-stable periodic structure of spin-transfer d.c. current can arise with certain channel step and step-to-radius ratio. The ferromagnetic resonance conditions determine spin-transfer current density. Independent nano-oscillators creating intersecting arrays of channels can phase-lock on sub-micrometer distance, which depends on multilayer geometry and applied fields. By decreasing the layer thickness, the number of channels affected by each independent d.c.-current source and their radius may be diminished. Phase-locking of multiple independent nano-oscillators can be used for enhancement of output power.

## 1. Introduction.

Electrical control of the electronic-spin degree of freedom is of basic interest for such spintronic applications as ferromagnetic and semiconductor devices contingent, as a rule, on injection of spin-polarized carriers and providing spin relaxation time and output power sufficient for device operation [1,2]. The theoretical predictions [3-6] and measurements [7-9] showed that the beyond-critical spin-polarized d.c. current in F-N-F multilayer causes resonance microwave excitation with the spin-precession frequency 5-40 GHz. In the recent experiments [10,11], two spin-transfer nano-oscillators phase-locked in proximity of 180-500 nm that creates exciting prospects of enhancement of output power, which is proportional to the square of the number of coherent oscillators. However, before such a device can be used on the nanoscale, phase-locking among many nano-oscillators with independent d.c.-current sources must be demonstrated. Practically, for critical current $I_c$ ~10 mA it's difficult to further increase density of independent d.c. sources, though phase-locking distance in the experiments was much more than electron spin-flipping distance 5-10 nm [5,6,12]. For real microwave bulk damping at such distance, obtained self-locking frequency band is explicable only for assumption of strong nonlinearity [13] that further complicates theoretical analysis necessary for parameter optimization. For resolving contradictions arising from the difference by the order of magnitude between phase-locking and spin-flipping distances, the real spatial structure of the spin-transfer current in the resonance regime should be considered – the problem addressed in the present submission.

In F-N-$F_{fix}$ multilayer (Fig.1, a), the electrons flow from free ferromagnetic F layer through normal N to fixed $F_{fix}$ layer in applied electric field $\vec{E}$. In F layer, electrons precess in external magnetic field $\vec{H}$ (not shown) around localized spin direction $\vec{S}_F$. The field $\vec{H}$ along with current $I \geq I_c$ determines precession frequency $\omega_F$, at that drive torque overcomes dumping at some critical current $I_c$, proportional to F layer thickness and determined by Gilbert bulk damping parameter $\alpha_F$ [3, 4, 14-16]. Spins of electrons pumped by electric field $\vec{E}$ to nonmagnetic N spacer, which separates F from $F_{fix}$, can be parallel or anti-parallel to the fixed $\vec{S}_{fix}$ direction characterizing $F_{fix}$ layer. At F-N interface magnon energy quanta $\hbar\omega_F$ are emitted. In the resonance regime this energy is equal to the difference in spin-up and spin-down Fermi levels in N (Fig. 1, b). The quanta $\hbar\omega_F$ are instantly absorbed that allows electrons to jump over the gap between Fermi levels, $\Delta\mu_N^{\uparrow\downarrow} = \mu_N^\uparrow - \mu_N^\downarrow$.

The spin distribution across N layer is determined by drift-diffusion mechanism. The diffusion model of spin distribution [17] was developed recently taking into account electric field effect [18-20]. The 1D quasi-steady-state approximation gives two diffusion lengths for injected spins [18] that could be attributed to the field reversing. The drift-diffusion model was generalized for three-component spin-polarization in electro-magnetic field with respect to crystallographic directions; cases of stationary distribution in direction perpendicular to the boundary [19] and of homogeneous relaxation [20] were reflected. Considering 1D electric-field distribution differential equation reveals that electron spin transport between, for example, two semiconductor regions with different doping levels, due to the charge density difference between layers, could lead to the charge redistribution and spin polarization amplification near their interface [21].

## 2. Model

Let's consider drift-diffusion partial differential equation (DDPDE) for 3D semi-infinite electron gas occupying half-space $x \geq 0$ in electric field $\vec{E}$ with an external source of spin gradient at boundary $x=0$ (see Fig. 1, a). DDPDE is derived from continuity equation and spin-current relations [21]. In the bulk of half-space, the electric field $\vec{E}$ ($-eE > 0$) is assumed to be invariable, uniform, and perpendicular to the boundary $x=0$. The right-hand part of DDPDE consists of diffusion and gradient polarization terms, field fluctuation and polarization attenuation terms. For F-N boundary between ferromagnet and normal metal, the boundary conditions can be expressed in the fifth term through the Dirac's $\delta$-function formalism:

$$\frac{\partial P}{D \partial t} = \Delta P + \frac{eE_{\nabla P}}{k_B T} \nabla P + \frac{e \nabla E}{k_B T} P - \frac{P}{D t_{sf}} + P'_x \delta(x), \qquad (1)$$

where, $P = (n_\uparrow - n_\downarrow)/n_e$ is a density polarization, defined as a superposition of spin-up and spin-down states, $n_e$ - charge density (fluctuations of which considered small in the bulk of N layer), $-e$ - the electron charge, $E_{\nabla P}$ - electric field intensity projection on $P$ gradient direction, $D = v_{drift} k_B T / eE$ - diffusion coefficient, $k_B$ - the Boltzman constant, $T$ - temperature, $t_{sf}$ - spin relaxation time, $P'_x = \partial P / \partial x$ - spin source gradient given at $x=0$. For the spin current in F-N-F$_{fix}$ junction, the boundary conditions in DDPDE (1) should be written for $x = 0$ and $x = d_N$:

$$P'_x(0) = P'_0, \qquad P'_x(d_N) = P'_{d_N}. \qquad (2)$$

Setting the boundary conditions for F-N-F$_{fix}$ junction in the resonance regime, one should distinguish between some averaged spin-current density $\bar{j}_F^{\uparrow\downarrow}$ correspondent to the beyond-critical charge current density in F layer, $-\bar{j}_F \geq -\bar{j}_{Fc} > 0$, and the local spin-current density $j_0^{\uparrow\downarrow} = P(0) j_0$ through F-N interface determined by angular-momentum conservation in electron-magnon interaction (here, $j_0 < 0$ is the local density of charge current perpendicular to the interface) [5, 6]. The electron gas is "pumped" through F-N interface most effectively (with zero interface resistance), if electrochemical potentials near spin-selective F-N interface match, i.e., the energy of an electron spin-flipping in N is equal to the energy of a magnon created in F:

$$\Delta \mu_N^{\uparrow\downarrow} - \hbar \omega_F = 0, \qquad (3)$$

$$\Delta \mu_N^{\uparrow\downarrow} = -2 j_0 P_{jF} \hbar k_x / e n_e, \qquad (4)$$

$$P(0) \approx P_{jF}, \qquad P(d_N) \approx P_{Ffix}. \qquad (5)$$

Here, $\Delta \mu_N^{\uparrow\downarrow}$ is a local spin-up(down) Fermi level shift caused by the spin current, $P_{jF}$, $P(0)$ - polarization at both sides of F-N interface, $P(d_N)$, $P_{Ffix}$ - polarization at both sides of N-F$_{fix}$ interface, $k_x$ - Fermi wave number, $\hbar = h/2\pi$ - reduced Planck constant.

The DDPDE (1) transforms to one without attenuation after introducing dimensionless diffusion coefficient $\bar{D}$ and dissipationless polarization function $p(x, y, z, t)$:

$$\overline{D} = \left(\frac{eE}{k_B T}\right)^2 D t_{sf},\tag{6}$$

$$P = p(x,y,z,t)\exp(-t/t_{sf}).\tag{7}$$

The equivalent diffusion equation (EDE) without drift-diffusion term for half-space $\xi \geq 0$ can be obtained by the coordinate transformation [22] and time transformation after introducing the characteristic electric field intensity length, $L_E$:

$$L_E = k_B T/(-eE),\tag{8}$$

$$\xi = \exp\left(\frac{x}{L_E}\right) - 1, \quad \psi = \frac{y}{L_E}\exp\left(\frac{x}{L_E}\right), \quad \zeta = \frac{z}{L_E}\exp\left(\frac{x}{L_E}\right),\tag{9}$$

$$\tau = \frac{t}{t_{sf}}\exp\left(\frac{x}{L_E}\right);\tag{10}$$

$$\frac{\partial p}{\overline{D}\partial \tau} = \left(\frac{\partial^2}{\partial \xi^2} + \frac{\partial^2}{\partial \psi^2} + \frac{\partial^2}{\partial \zeta^2}\right)p + p'_{j\xi}(\xi,\psi,\zeta,\tau)\delta(\xi),\tag{11}$$

$$p'_{j\xi}(\xi,\psi,\zeta,\tau) = P'_{jx}\frac{k_B T}{-eE(1+\xi)}\exp\left(\frac{\tau}{(1+\xi)^2}\right).\tag{12}$$

For initial condition $p(\xi,\psi,\zeta,0) = 0$, the partial solution of the EDE (11), (12) can be obtained in the integral form [23]:

$$p = \frac{1}{\sqrt{\pi}}\int_0^\tau p'_{j\xi}(0,\psi,\zeta,\tau')\exp\left(-\frac{\xi^2}{4\overline{D}(\tau-\tau')}\right)\frac{\overline{D}d\tau'}{\sqrt{\overline{D}(\tau-\tau')}}.\tag{13}$$

Correspondingly, in $(x, y, z, t)$ coordinate system the integral (13) can be expressed as:

$$P = \frac{\sqrt{Dt_{sf}}}{\sqrt{\pi}}\int_0^{\frac{t}{t_{sf}}\exp\left(\frac{-2eE}{k_B T}x\right)} P'_{jx}\exp\left(\tau' - \frac{\tau}{e^{\frac{-2eE}{k_B T}x}} - \frac{\left(e^{\frac{-eE}{k_B T}x} - 1\right)^2}{4\overline{D}(\tau-\tau')}\right)\frac{d\tau'}{\sqrt{\tau-\tau'}}.\tag{14}$$

### 3. Results and discussion.

The quasi-steady state, $\partial P/\partial t = 0$, takes place at time limit $t/t_{sf} > 1$. Analysis of DDPDE partial solution for zero initial polarization and for boundary conditions (2) shows that the similar quasi-stationary condition exists already at moment $t/t_{sf} = 0.15$. In equivalent coordinate system, a quasi-steady state corresponds to constant polarization gradient, at that an equivalent cross-section area of current channel is constant across N layer. Due to the character of coordinate transformation (9), in real-space coordinates, this cross-section is exponentially narrowing along $x$ (such channels are shown in Fig. 2).

If N layer is thinner than mean-free path, $d_N < L_{sf} \approx 2\sqrt{Dt_{sf}}$, the gradient is exponentially dependent on the characteristic electric field intensity length, $L_E$:

$$P'_x = P'_0 \exp(x/L_E).  \quad (15)$$

Complying with the quasi-steady-state relationship (15), the density polarization is distributed across N layer exponentially (Fig. 2, b). Taking into account spin-current relations for sufficient electric field intensity [21], for boundary conditions (3)-(5) one can obtain:

$$P_{Ffix} - P_{jF} = P'_0 L_E (\exp(d_N/L_E) - 1), \quad L_E = \left(\frac{P'_0}{P_{jF}} - \frac{j_0}{en_e D}\right)^{-1},  \quad (16)$$

$$P_{jF} = -\frac{\omega_F en_e}{2 j_0 k_x}.  \quad (17)$$

Expression (16) can be simplified for the case of the quasi-steady spin-conductivity regime:

$$P'_0 2k_x/\omega_F \gg D^{-1}, \qquad L_E \approx P_{jF}/P'_0.  \quad (18)$$

The critical charge current is proportional to ferromagnetic-resonance frequency, $\omega_F$, determined by magnetic field in F and, due to the Gilbert damping, to F layer thickness, $d_F$ [14]. The current density at F-N boundary, $j_0$, is determined by $\omega_F$ according to (17). Therefore, the affected area is proportional to $d_F$.

In transition to the resonance regime, the electron-magnon interactions at inhomogeneous F-N interface can be associated with some actual fluctuations of interface resistance. During the transition period after free F layer repolarization, due to the local charge density difference between layers, the electric field and polarization fluctuations at the boundary $x = 0$ take place with magnitude of order of their stationary level [21] (Fig. 2, a). Out of all possible fluctuation modes, the prevailing spatial mode corresponds to the case when the gradient term in DDPDE (1) becomes large enough to overcome the attenuation term:

$$e\nabla E/k_B T \geq (Dt_{sf})^{-1}.  \quad (19)$$

Like in the fractal growth during a phase transition, fluctuations along the boundary commonly may result in developing a periodic structure with spatial characteristics, which, in the quasi-steady-state limit, are determined by condition of Laplacian field potential energy minimization in transformed coordinates [22]. According to (19), the periodic polarized current structure can be initiated by spatial field fluctuations (Fig. 2, a) if the next condition for their characteristic size $2r$ is satisfied:

$$r \leq \pi L_{sf}^2/4L_E.  \quad (20)$$

The equilibrium condition for 2D spin-transfer current lattice, where each current channel is surrounded by eight others with period $\lambda$ (Fig. 2, b, c), can be determined as a minimum of a vector potential [24] for an outer ring of a current-channel cross-section with radius $r$:

$$\frac{\partial}{\partial r}\left(\frac{2}{r} + 6\left(1.74\left(\frac{1}{\lambda - r} + \frac{1}{\lambda + r}\right) + \frac{2}{\sqrt{\lambda^2 + r^2}}\right)\right) = 0,  \quad (21)$$

$$\lambda / r \approx 2.8 \,. \tag{22}$$

The optimum step-to-radius ratio (22) is characteristic for the fractal structures evolving at relatively small external gradient, such as the secondary dendrites at crystallization [22]. For the case of thick N layer (small $L_E$) more complex fractal structure evolves. The average-local current density relation for thin N layer is determined by an area ratio:

$$\bar{j}_0 / j_0 = \pi r^2 / \lambda^2 \approx 0.40 \,. \tag{23}$$

For F-N-F$_{fix}$ multilayer parameters, $P_{Ffix} = 0.3$, $P_{jF} \approx 0.2$, $d_N = d_F = 5$ nm, $k_x = 13.6$ nm$^{-1}$, $\omega_F = 6.8$ GHz, $D = 1.6 \cdot 10^{-4}$ m²/s, $L_{sf} = 10$ nm, the characteristic electric field intensity length is $L_E = 12.3$ nm, the current-channel radius, $r = 6.4$ nm and their period, $\lambda = 18.0$ nm. According to (17), (23), the local and average current densities on F-N boundary are $j_0 = 1.07 \cdot 10^{11}$ A/m², $\bar{j}_0 = \bar{j}_F = 4.27 \cdot 10^{10}$ A/m². For the resonance current 9mA, the affected area should be $2.1 \cdot 10^5$ nm², which includes ~655 current nano-channels, with characteristic radius approx. $R_1 = R_2 = 260$ nm (Fig. 3), that corresponds to results for phase-locking distance between independent nano-oscillators [10, 11]. Because the critical current and affected area are proportional to $d_F$, the phase-locking distance is proportional to $d_F^{0.5}$. The channel radius according to (20) is inversely proportional to $L_E$ and, correspondingly, to $d_N$.

A device for the experimental study can be built on the base of the nano-contact GMR spin-valve [10, 11], for example: Ta 5 nm/Cu 50 nm base electrode, Co$_{90}$Fe$_{10}$ 20 nm fixed magnetic (F$_{fix}$) layer, Cu 5 nm (N) spacer, Ni$_{80}$Fe$_{20}$ free magnetic (F) layer, Cu 1.5 nm/Au 2.5 nm cap. The channel point contacts should be fabricated with step multiple of $\lambda = 18.0$ nm as extreme ultraviolet lithography allows such resolution. The current at each contact point and the distance between them should be determined by F layer thickness that can be smoothed by gas-cluster ion-beam technology to 1 nm. In the ideal device, decreasing F, N layer thicknesses to single atomic layers should allow not only drastically increase the density of coherent nano-oscillators, but also use their interaction for the quantum computing.

**4. Conclusions**

Conditions of ferromagnetic resonance in F-N-F multilayer devices determine the current density at F-N interface, at that the critical current and affected area for each independent spin-polarized d.c. current source is proportional to F-layer thickness. In transition to the resonance regime, the spatial instability modes arise and out of all possible fluctuation modes, the prevailing quasi-stable spatial mode corresponds to the criterion of minimum vector potential. The periodic spatial current structure represents an array of channels with nanometer-size step and step-to-radius ratio 2.8 corresponding to low field intensity. By decreasing F, N layer thickness the radius and number of channels affected by each independent d.c.-current source may be diminished. Phase-locking of multiple independent nano-oscillators can be used for enhancement of output power.

**Acknowledgements**

The author thanks Dr. Isaak Zaritsky and Dr. Chih-Ling (Fabian) Lee of Veeco Instruments Inc. for helpful discussions.

**Figures**

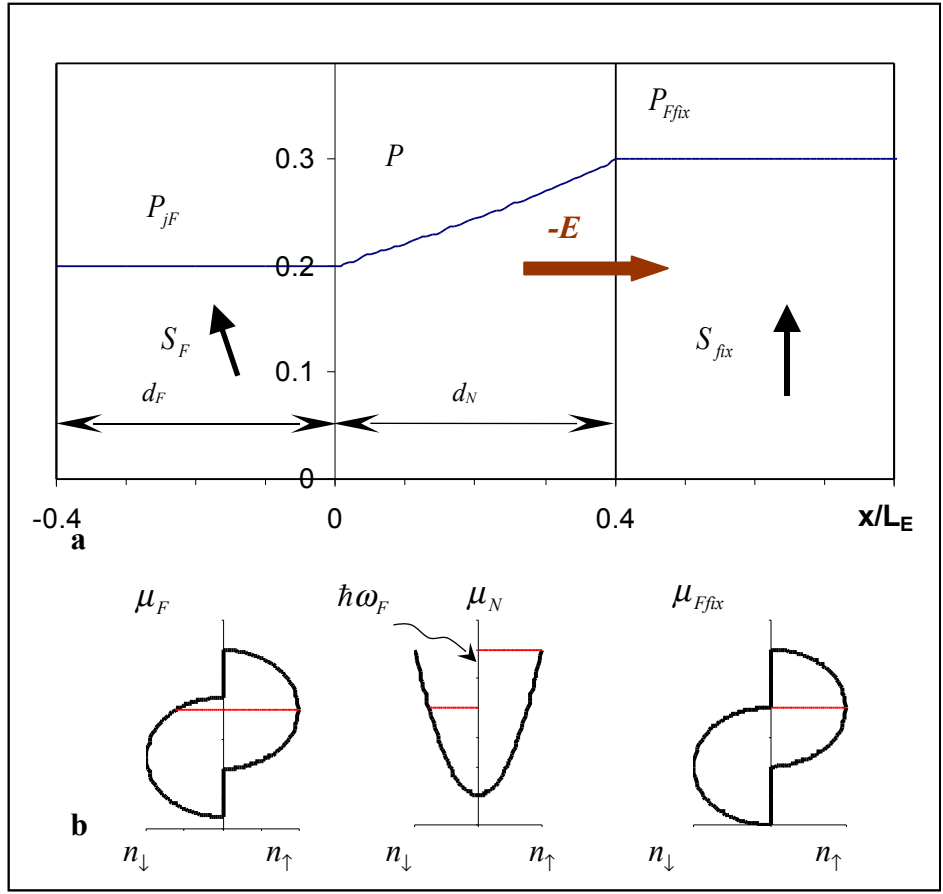

Fig. 1. Overview of F-N-F device in the resonance regime. a, Polarization distribution in the "free" ferromagnetic F layer, normal spacer N, and "fixed" $F_{fix}$ layer. The shown free-layer polarization $P_{jF}$ corresponds to the magnetic spin $\vec{S}_F$ projection on the fixed-layer spin $\vec{S}_{fix}$ direction with polarization $P_{Ffix}$. Polarization $P$ in N layer corresponds to zero F-N interface resistance. b, Spin-resolved densities of states across F-N-$F_{fix}$ junction. Unequally filled spin-up(down) levels $\mu_N$ in the density of states depict spin-resolved electrochemical potential difference of the electron flow through F-N interface; this difference corresponds in the resonance regime to the magnon energy quant $\hbar\omega_F$.

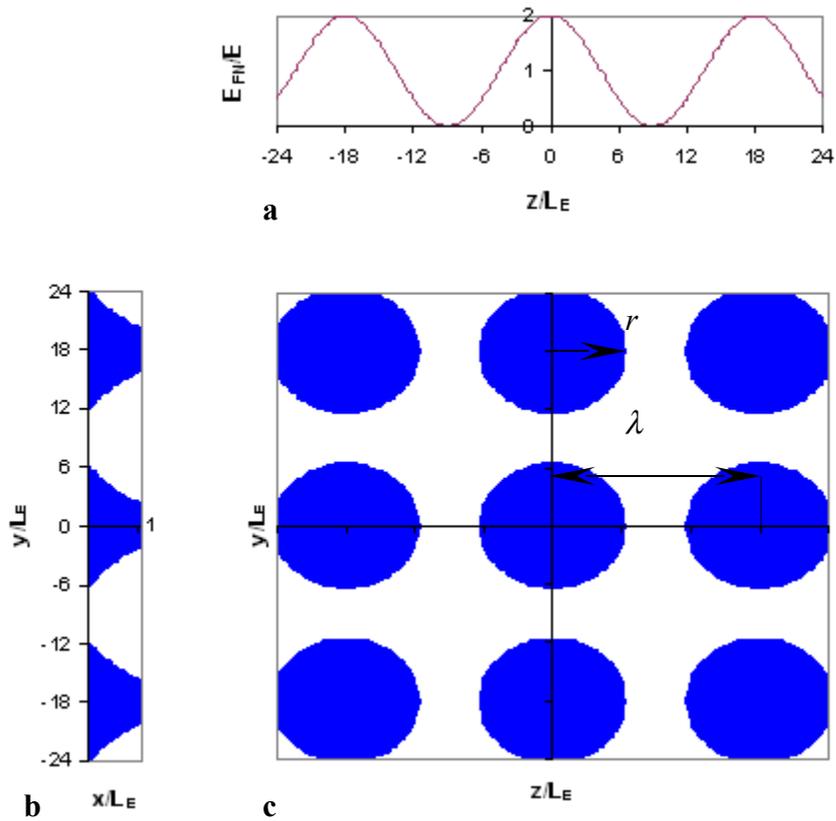

Fig. 2. Self-organized structure of spin-transfer dc current channels. a, The prevailing initial field intensity spatial fluctuation mode at F-N interface. Field intensity magnitude $E_{FN}$ is related to the stationary value $E$. b, The spin-transfer d.c.-current channel exponential profile across N layer. c, The periodic structure of current channels. $r$ - quasi-stable channel radius; $\lambda$ - channel step.

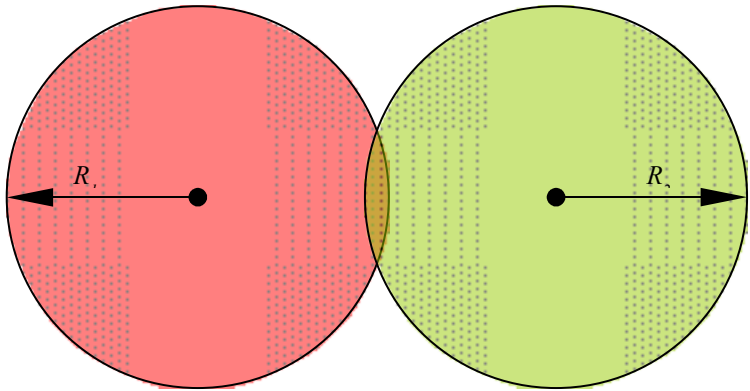

Fig. 3. Intersecting arrays of coherent oscillators. Two arrays of spin-transfer dc-current channels generate phase-locked coherent oscillations.